\documentclass[%
 reprint,
 superscriptaddress,
nofootinbib,
 amsmath,amssymb,
 aps,
 prd,
floatfix,
twocolumn,
notitlepage
]{revtex4-1}
\usepackage{graphicx}
\usepackage{dcolumn}
\usepackage{subfigure}
\usepackage{bm}
\usepackage{breakurl}
\usepackage{enumitem}
\usepackage{xcolor}
\usepackage[normalem]{ulem}
\usepackage{multirow}

\usepackage[multidot]{grffile}
\newcommand{\bea}{\begin{eqnarray}}
\newcommand{\eea}{\end{eqnarray}}

\begin{document}

\title{Local topological moves determine global diffusion properties of \\hyperbolic higher-order networks}

\author{Ana P. Mill\'an}
\affiliation{
Amsterdam UMC, Vrije Universiteit Amsterdam, Department of Clinical Neurophysiology and MEG Center, Amsterdam Neuroscience, De Boelelaan 1117, Amsterdam, The Netherlands}
\author{Reza Ghorbanchian}
\affiliation{School of Mathematical Sciences, Queen Mary University of London, Mile End Road, E1 4NS, London, United Kingdom}
\author{Nicol\`o Defenu} 
\affiliation{Institute for Theoretical Physics, ETH Z\"urich
Wolfgang-Pauli-Str. 27, 8093 Zurich, Switzerland}
\author{Federico Battiston}
\affiliation{Department of Network and Data Science, Central European University, 1100 Vienna, Austria}
\author{Ginestra Bianconi}
\affiliation{School of Mathematical Sciences, Queen Mary University of London, Mile End Road, E1 4NS, London, United Kingdom}
\affiliation{The Alan Turing Institute,  British Library, 96 Euston Road,  NW1 2DB, London, United Kingdom}

\begin{abstract}
From social interactions to the human brain, higher-order networks are key to describe the underlying network geometry and topology of many complex systems. 
While it is well known that network structure strongly affects its function, the role that network topology and geometry has on the emerging dynamical properties of higher-order networks is yet to be clarified.
In this perspective, the spectral dimension plays a key role since it determines the effective dimension for diffusion processes on a network.
Despite its relevance, a theoretical understanding of which mechanisms lead to a finite spectral dimension, and how this can be controlled, represents nowadays still a challenge and is the object of intense research. 
Here we introduce two non-equilibrium models of hyperbolic  higher-order networks and we characterize their network topology and geometry by investigating  the interwined appearance of small-world behavior, $\delta$-hyperbolicity and community structure. 
We show that different topological moves determining the non-equilibrium growth of the higher-order hyperbolic network models induce  tunable values of the spectral dimension, showing a rich phenomenology which is not displayed in random graph ensembles. 
In particular, we observe that, if the topological moves used to construct the higher-order network increase the area$/$volume ratio, the spectral dimension continuously decreases, while the opposite effect is observed if the topological moves decrease the area$/$volume ratio.
Our work reveals a new link between the geometry of a network and its diffusion properties, contributing to a better understanding of the complex interplay between network structure and dynamics.
\end{abstract}

\maketitle

\section{Introduction} 

Higher-order networks are  generalized network structures that capture the many-body interaction of complex systems\cite{battiston2020networks, torres2020and, bianconi2015interdisciplinary, salnikov2018simplicial, giusti2016two}. In recent years, they have become increasingly popular to represent different types of data beyond the framework of pairwise interactions, including the human brain \cite{petri2014homological, reimann2017cliques,  tadic2019functional, fernando2021tutorial}, 
social interacting systems \cite{petri2018simplicial, iacopini2019simplicial, landry2020effect, st2021bursty, jhun2019simplicial,de2020social,
carletti2020random, matamalas2020abrupt, cencetti2021temporal}, 
financial networks \cite{tumminello2005tool, massara2016network}, and complex materials \cite{bassett2012influence, vsuvakov2018hidden,tadic1}.  Interestingly, several studies on synchronization, diffusion, epidemic spreading and evolutionary dynamics 
have shown that taking into account the higher-order organization of networks can lead to emergent  behavior remarkably different from that of graphs, where interactions are limited to groups of two nodes only \cite{millan2020explosive, millan2018complex, millan2019synchronization, ghorbanchian2020higher, skardal2019abrupt, millan2020explosive, lucas2020multiorder,   mulas2020coupled, alvarez2021evolutionary, salova2021cluster}. 
Therefore, establishing the relation between the structure and the dynamics of higher-order networks is currently a field of intense research activity.

A powerful tool to characterize higher-order network data is Topological Data Analysis, which provides a general mathematical and computational framework to analyze data from their topological shape \cite{petri2014homological, otter2017roadmap, kartun2019beyond, lee2020homological, bobrowski2020homological}. 
In a number of cases, these analyses have been able to extract  relevant information from real networks that cannot be detected by more traditional Network Science metrics. Moreover, topology can directly affect higher-order network dynamics by determining the evolution of topological signals, i.e. dynamical signals defined not only on nodes but also on links, triangles, and other higher-order structures  \cite{taylor2015topological, barbarossa2020topological, millan2020explosive, torres2020simplicial}. 
A natural way to represent higher-order networks is by simplicial complexes, which differently from graphs are not only formed by nodes and links, but also include triangles, tetrahedra and so on. 
Together with cell complexes -- which allow as building blocks also other regular polytopes such as hypercubes and orthoplexes -- simplicial complexes describe discrete topological spaces. 
Therefore, modelling higher-order networks with simplicial and cell complexes is often the first step for conducting a topological investigation of higher-order networks.

Models of dynamically evolving simplicial and cell complexes built by simple local rules can affect the higher-order network topology by causing the emergence of a non-trivial meso-scale topological organization. 
For instance, models of simplicial complexes implementing triadic closure  \cite{bianconi2014triadic}, or constructed by gluing together simplices through the iteration of simple topological moves  \cite{bianconi2017emergent, mulder2018network, vsuvakov2018hidden, tadic1} were found to generate higher-order networks with emergent community structure.
Moreover, simplicial complexes and cell complexes have also an intrinsic geometrical nature, and for this reason represent an ideal setting to investigate the properties of emergent hyperbolic network geometry  in complex systems \cite{bianconi2015interdisciplinary, wu2015emergent, bianconi2016network, bianconi2017emergent, vsuvakov2018hidden, tadic1}.
In particular, models of emergent hyperbolic network geometry reveal the fundamental rules responsible for  the wide-spread occurrence of hyperbolicity in real network  datasets and in the structure of knowledge graphs  \cite{gromov1987hyperbolic, albert2014topological, narayan2011large, chami2019hyperbolic,  chami2020low}.

A characteristic feature of emergent geometries is revealed by the spectral properties of their network skeleton, i.e. the network that is generated from the higher-order system only retaining the pairwise interactions. 
In particular, a fundamental indication that the higher-order networks have a characteristic geometrical nature is associated with the emergence of a finite spectral dimension of their graph Laplacian. 
Broadly speaking, the spectral dimension $d_S$ of the graph Laplacian of a network indicates the dimension of the network as perceived by a random walker crawling on the network. 
As such, on a regular Euclidean lattice the spectral dimension $d_S$ coincides with the dimension of the lattice $d$. 

Until recently, it was believed that the heterogeneous degree distributions and the Fiedler eigenvalue  of the graph Laplacian were the main structural determinants for dynamical processes in networked systems. 
These considerations, however, reveal how the investigation of higher-order networks has recently transformed the way in which we look at the classical problem of the interplay between structure and dynamics on complex networks  \cite{Dorogovtsev2008}. 
From the study of higher-order networks, it is now becoming clear that both network topology and network geometry can affect dynamics in unexpected ways that go well beyond previous beliefs. 
In particular, the spectral dimension constitutes a fundamental quantity to capture how geometry affects dynamics.  
For instance, it is is known to characterize  the return time of random walkers\, \cite{Burioni1996, masuda2017random}, the stability of synchronised states\, \cite{millan2018complex, millan2019synchronization}, universal critical phenomena\, \cite{Joyce1966, Bradde2010, aygun2011spectral,  Defenu2015, Defenu:2017dc, Gori2017, millan2020complex}, quantum diffusion\, \cite{mulken2011continuous, nokkala} or the universal properties of different quantum gravity approaches\, \cite{ambjorn2005spectral, benedetti2009fractal, astrid}.
Remarkably, the value of the spectral dimension is not  universal in  networks but can vary significantly from network to network. 
Moreover, in random graphs and expanders the spectral dimension is not even defined as their  highly non-local nature of their connections have the effect of introducing a spectral gap in the spectrum of the graph Laplacian.

Most of the models displaying emergent spectral dimension evolve by the iteration of simple topological moves that are also responsible for a  non-trivial community structure. 
Moreover, in the existing models of emergent hyperbolic geometry that display a finite spectral dimension, including the model \textit{Network Geometry with Flavor} \cite{mulder2018network, millan2018complex, millan2019synchronization, tadic1}, the dependence of the spectral dimension on the model parameters has yet to be clarified. 
Indeed, how a finite spectral dimension emerges is still an open problem.  
Which are the general microscopic mechanisms giving rise to a finite spectral dimension? What is its relation to  hyperbolicity of the associated emergent geometries?

In this work we address these these questions by systematically investigating the relation between the simple topological moves determining the local evolution of higher-order networks and their meso-scale and global properties. 
Our results are based on the analyses of two distinct classes of models -- the Short-Range Triadic Closure (STC) model and the Network Geometry with Flavor (NGF) model- that generalize previous models enforcing triadic closure  \cite{bianconi2014triadic} and displaying emergent hyperbolic geometry  \cite{bianconi2017emergent} respectively.
We illustrate how different local topological moves can lead to important differences in the large-scale structural and dynamical properties of the higher-order networks. 
Specifically, we show how different topological moves can be used to increase or decrease the value of the spectral dimension and the modularity of hyperbolic higher-order networks. 
Interestingly, we directly link the emergence of a smaller spectral dimension to topological moves that enforce a larger ratio between the area and the volume of the considered emergent hyperbolic geometries.

The paper is structured as follows. 
In Sec. II we define the two classes of models: the STC and the NGF model. 
In Sec. we discuss the topological properties of these models, including a detailed discussion of the role of the topological moves on the evolution and emergence of non-trivial community structure in both models. 
In Sec. IV we characterize the emergent hyperbolic geometry of the STC and the NGF model. 
In Sec. V we show how the emergent spectral dimension of both models is modulated by the choice of topological moves adopted for the evolution of the higher-order network models and the implications for diffusion dynamics. 
Finally, in Sec. VI we provide some concluding remarks.

\section{Higher-order network models and their underlying network skeleton}

\subsection{Mathematical definition of cell complexes and network skeleton}

Higher-order networks allow to represent networked systems which are not limited to only pairwise interactions. 
A common way to describe such structures is to introduce new higher-order building blocks known as \textit{simplices}.
A $d$-dimensional simplex is formed by $d+1$ nodes, each one interacting with all the other ones. 
Thus, a $d$-dimensional simplex is a  node when $d=0$, a  link when $d=1$, a triangle when $d=2$, a tetrahedron when $d=3$, and so on. 
The underlying network skeleton of a $d$-simplex, i.e. the network retaining only the pairwise interactions between the nodes, is a $(d+1)$-clique. 
\textit{Simplicial complexes} -- a collection of simplices which respect a particular inclusion rule of their lower order faces -- provide a good representation of higher-order networks. 
The \textit{facets} of a simplicial complex are the simplices that are not faces of any other simplices of the simplicial complex. 
The \textit{dimension} of a simplicial complex is the maximum dimension of its facets.
The \textit{skeleton} of a simplicial complex is the network constructed from the simplicial complex retaining only the information about its nodes and links.

In a number of real-world scenarios, however, higher-order networks may be constructed by buildind blocks which are more loosely connected than simplices. 
For instance, a protein interaction network is formed by a set of proteins that might have a complex interaction pattern involving more than two agents, but might not be binding to each other in an all-to-all small subgraphs.
These generalised building blocks are know as \textit{cells}. 
Mathematically, a $d$-dimensional  cell is  a $d$-dimensional convex polytope, i.e.  a topological space homeomorphic to a $d$-dimensional open ball. 
Therefore, $0$-dimensional cells are nodes and $1$-dimensional cells are links, and therefore do not differ from $0$-dimensional  and $1$-dimensional simplices. 
However, differences originate in higher dimensions. 
For instance, $2$-dimensional cells include $m$-polygons such as triangles ($2$-dimensional simplices), but also squares, pentagons, etc. 
Similarly, $3$-dimensional cells include the Platonic solids, such as tethrahedra ($3$-dimensional simplices), cubes, octahedra, dodecahedra, and icosahedra. 
Interestingly, whereas in dimension $d=4$ there are more regular polytopes than in dimension $d=3$ (being $6$), for any dimension $d>4$ there are only three types of regular (convex) polytopes: the simplex, the hypercube and the orthoplex.

Similarly to simplicial complexes, cells may be aggregated into a cell complex.
In particular, a cell complex $\hat{\mathcal K}$ has the following two properties:
\begin{itemize}
\item[(a)]it is  formed by a set of cells that is closure-finite, meaning that every cell is covered by a finite union of open cells;
\item[(b)] given two cells of the cell complex $\alpha\in \hat{\mathcal K}$ and $\alpha^{\prime}\in \hat{\mathcal K}$ then either their intersection belongs to the cell complex, i.e. $\alpha\cap \alpha^{\prime}\in \hat{\mathcal K}$ or their intersection is a null set, i.e. $\alpha\cap \alpha^{\prime}=\emptyset$.
\end{itemize}

The dimension $d$  of a cell-complex is the maximum dimension of its cells. 
Therefore, a $d$-dimensional Euclidean square lattice can be seen as a cell-complex of dimension $d$ as it is formed by  unit cells which are hypercubes of dimension $d$.
Similarly to simplicial complexes, the skeleton of a cell-complex is the network generated from the cell complex only retaining the pairwise interactions between its nodes.

\subsection{The building blocks of the proposed higher-order models}

In the following we consider two non-equilibrium models of higher-order networks. 
Each model is characterized by the non-equilibrium dynamics describing the higher-order network growth. 
At each time step one or more nodes are added and connected to the rest of the network through a specific higher-order interaction, representing the building blocks of our networked structures, and taken to be either  $d$-dimensional simplices or $d$-dimensional orthoplexes.
The $d$-dimensional simplexes have been defined in the previous section as formed by $d+1$ nodes each one interacting with all the other ones. 
The $d$-dimensional orthoplex is a regular polytope with $2d$ nodes and $2^d$ faces formed by $(d-1)$- dimensional simplices.
For instance, in  $d=2$ the orthoplex is a square having $4$ nodes and $4$ links, for $d=3$ the orthoplex is a bi-pyramid with a square basis having $6$ nodes and $8$  triangular faces.
In general, the network skeleton of a collection of a $d$-dimensional orthoplex is more sparse than the one of $d$-dimensional simplices, which is a fully connected clique.

Since both $d$-simplices and $d$-orthoplexes admit as $(d-1)$-faces  exclusively  $(d-1)$-dimensional simplices, orthoplexes and simplices can be easily glued to each other and combined as coexisting building blocks of a higher-order network (also called cell complex).
For instance, in our higher-order network models defined in  dimension $d=2$, we will combine triangles and squares glued along their links in higher-order discrete architectures. 
However, $d$-simplices and $d$-orthoplexes can be glued to each other according to different topological moves. In order to reveal the macroscopic consequences of the choice of different local moves, here we consider two specific models:  the Short-Range Triadic Closure (STC) model,  and a new variation  of the  Network Geometry with Flavour  (NGF)  model proposed in Refs.\, \cite{bianconi2016network, mulder2018network}.

\begin{figure*}
\centering
\includegraphics[width=1\textwidth]{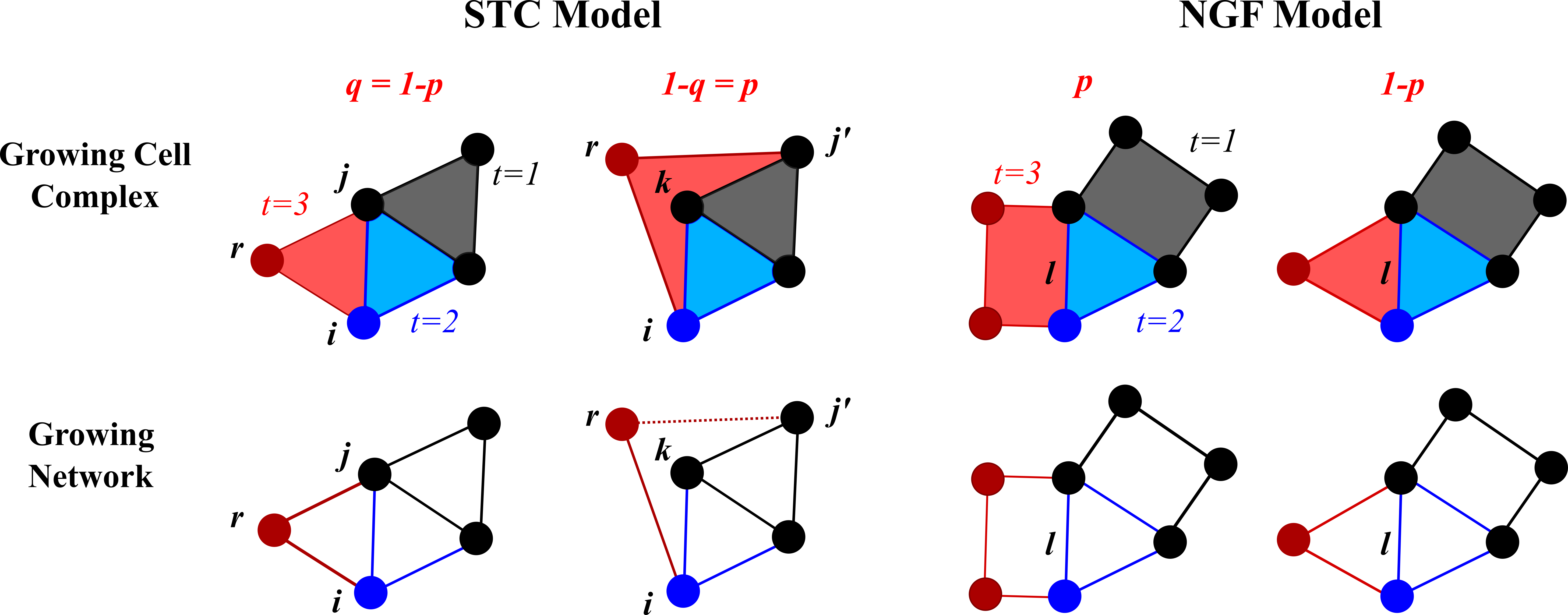}
\caption{ 
Illustrative sketch of the different topological moves used  to grow higher-order networks (top panels) for the STC (with $m=2$) and NGF (with $d=2$) models, and their corresponding network skeletons (bottom panels). 
In the case of the STC model, at each time an existing node $i$ is randomly selected. 
Then, with probability $q$ a first neighbour of node $i$, let say node $j$, is randomly chosen, and a new node $r$ is linked to both $i$ and $j$, thus forming a triangle.
By contrast, with probability $1-q$ a second neighbour $j'$ of node $i$ is selected  and the new node $r$ is linked to both $i$ and $j'$, closing a square. 
The fourth node $k$ conforming the square is randomly selected among the shared neighbours of $i$ and $j'$.
For the NGF model, at each time-step an existing link $l$ is randomly selected according to Eq. ($\ref{eq:NGF}$) and either a square (with probably $p$) or a triangle (with probability $1-p$) is added to the cell complex. 
 \label{fig:fig1}}
\end{figure*}

\subsection{Short-Range Triadic Closure (STC) model}

The Short-Range Triadic Closure  (STC) model is a higher-order network model that generalizes triadic closure models considering not only the introduction of triangles but also of squares. 
The model is defined as follows. 
Initially (at $t=1$), the network is formed by an $m$-dimensional simplex.
At each time step $t>1$, a new node $r$ is added to the network and connected to the rest of the higher-order network by $m$ links and by $m-1$ higher-order interactions.
The first link of the new node is connected to a  randomly selected node $i$. 
The remaining $m-1$ links are chosen in such a way to close triangles with the first link $(r,i)$ with probability $q$, and to close squares with probability $1-q$. 
If the first event occurs, the new link is connected to a random neighbour $j$ of node $i$, forming the triangle $(r,i,j)$. 
On the contrary, if the second event occurs, the new link is connected to a random second neighbour $j'$ of node $i$, and a fourth node $k$ is selected among the common neighbours of $i$ and $j'$, and the square $(r,i,k,j')$ is formed.  
If $q=1$ the higher-order network grows exclusively by the addition of triangles, whereas for $q=0$ it grows by the addition of squares.
Therefore, the  STC model generates $(d=2)$-dimensional cell complexes for every value of $m\geq 2$. 
The process and the underlying network skeleton are illustrated in Fig. $\ref{fig:fig1}$ for $m=2$.

A long-range version of the model, without the limitation to connecting only to $1$\textsuperscript{st} or $2$\textsuperscript{nd} neighbours, was extensively studied in  \cite{bianconi2014triadic}. 
It was shown to have a heavy-tailed degree distribution, short paths, a strong community structure and high clustering (for $q>0$). 
The STC dynamics is only driven by an effective {\em sublinear} preferential attachment, i.e. new links are effectively attached to a generic node $i$ with a probability proportional to $k_i^{\theta}$ with $\theta<1$, however as $m$ increases the exponent $\theta$ approaches one, leading to broader degree distribution.

\subsection{ The Network Geometry with Flavor (NGF) model}

The second model that we  consider is a variation of the  Network Geometry with Flavor (NGF) model proposed for cell complexes  in Refs.  \cite{bianconi2016network,  bianconi2017emergent}.
In its original formulation, the NGF cell complexes evolved by the subsequent addition of a single type of regular polytope (i.e. all the polytopes of the cell complexes are the same). 
Here we generalize this modelling framework by allowing the cell complex to be formed by different types of $d$-dimensional polytopes.
In particular, the present version of the  NGF model generates $d$-dimensional higher-order networks  by subsequently gluing together $d$-dimensional simplices, or $d$-dimensional orthoplexes, along their $(d-1)$-faces. Each $(d-1)$-face $\alpha$ of the higher-order network is characterized by its incidence number $n_\alpha$, indicating the number of $d$-dimensional polytopes incident to the face, minus $1$. 

The higher-order network is constructed as follows. 
Initially, (at $t=1$) the NGF is formed by a single  $d$-dimensional orthoplex. 
At every subsequent time step $t>1$, a new $d$-dimensional polytope is glued to a $(d-1)$-face $\alpha$.
The new polytope is a $d$-dimensional orthoplex with probability $p$ and a $d$-dimensional simplex with probability $1-p$. 
The face $\alpha$ to which the new polytope is attached is chosen with  probability 
\begin{equation}\label{eq:NGF}
\Pi_\alpha = \frac{1+sn_\alpha}{\sum_{\alpha'} (1+sn_{\alpha'})},
\end{equation}
where $s\in \{-1,0,1\}$ is a parameter of the NGF model called  {\em flavor}.
For $p=0$, if we neglect the initial condition, the NGF is formed exclusively by $d$-dimensional simplices and reduces to the model treated in Ref.  \cite{bianconi2016network}. 
For $p=1$, the NGF is formed exclusively by $d$-dimensional orthoplexes, and  this limit has been studied in detail in Ref. \cite{mulder2018network}.

In general, NGFs comprise of $d$-dimensional cell complexes. 
In the following, we will often refer to their network skeleton as the underlying NGF network. 
NGF networks display interesting combinatorial  properties that have been well characterized in the case $p=0$  \cite{bianconi2016network} and $p=1$  \cite{mulder2018network}: the degree distribution is scale-free for a wide range of values of $s$ and $d$, and the resulting networks are small-world and have an infinite Hausdorff dimension. 
Despite the small-world property, NGFs have a finite spectral dimension together with a strong hierarchical-modular community structure  \cite{mulder2018network, millan2019synchronization}. 
In this work we fix the value of the flavor of  the NGF model to $s=0$, leading to power-law networks as the result of an effective preferential attachment mechanism for any dimension $d\geq 2$. 
A schematic illustration of this model is also represented in Fig. $\ref{fig:fig1}$ for $d=2$.

\section{Topological properties of the STC and the NGF models}

The skeleton of the STC and NGF cell complexes -- made up by considering only the nodes and links -- makes up the STC and NGF networks, whose properties can be evaluated through standard network science metrics. 
According to such type of analyses, the two models can be considered remarkably similar. 
They both lead to networks with high clustering, heterogeneous degree distributions, short paths and a strong community structure.
For instance,  we note the case  of $m=2$ (STC model) and $d=2$ (NGF model). 
For $q=1$ and $p=0$, the two networks are made up by gluing triangles together. Then, as $q$ decreases (STC) or $p$ increases (NGF), triangles are substituted by squares. 
Thus, in order to compare the two models in the following, we define the complementary control parameter $p=1-q$ for the STC model, so that increasing $p$ leads in both models to a decrease in the number of triangles (or, more generally, simpleces) and an increase in the number of squares (or orthoplexes) conforming the high-order network. 

Despite the aforementioned similarities from the network science perspective, the two models are constructed by adopting different topological moves, as we go on to show in the following subsections.

\subsection{Topological moves for the STC and NGF model}

The most striking  difference between the STC and the NGF cell complexes is that, while the STC cell complex is $d=2$ dimensional for every value of $m$, the NGF cell complex has varying dimension $d\geq 2$ given by the dimensionality of its building blocks. Besides, even if we limit our considerations only to STC and NGF cell complexes in dimension $d=2$, the two models differ by the dynamical rules used for their generation. 
These rules, determining the way in which simplices and orthoplexes are added to the cell complex, are called in topology {\em topological moves}. 

Both models share a remarkable feature, which is that the topological moves leading the evolution of the cell complexes do not change the topological invariants of the cell complexes.
In fact, simplices and orthoplexes are added to the cell complex in such a way that neither the Betti numbers  (with the only non-zero Betti number being $\beta_0=1$) nor the Euler characteristic of the cell complex $\chi$ change \cite{topo1,topo2}.
This latter property can be easily checked for both the STC  and the NGF models in $d=2$ using the well known definition of the Euler characteristic as alternating sum of the number $s_k$ of $k$-dimensional cells of the cell complex, with $0\leq k\leq d$  \cite{topo1,topo2}
(leading,  for $d=2$, to the famous expression $\chi=V-E+F$  where $V$, $E$ and $F$ are respectively the number of nodes, links and $2$-dimensional cells of the cell complex). 
For the STC model with $m=2$, a single $2$-dimensional cell is added at each time, which can be either a triangle glued to an existing link, or a square glued to two existing links. 
In both cases,  the cell complex increases by one node, $\Delta V=1$, two links, $\Delta E=2$, and one triangle, $\Delta F=1$, so the Euler characteristic changes by 
\bea
\Delta \chi=\Delta V-\Delta E+\Delta F=1-2+1=0.
\eea
For  the STC model with $m>2$, at each time, $m-1$ cells of dimension $d=2$ are added to the cell complex.
The first cell is glued to a link as for the case $m=2$, leading to $\Delta \chi=0$. 
The subsequent cells share a link with the first cell (the one between the new node and the randomly chosen node) and  add a new link, $\Delta E=1$, and a $2$-dimensional cell, $\Delta F=1$, to the cell complex, where the $2$-dimensional cell can be either a triangle or a square.
Therefore, none of the subsequently added cells change the Euler characteristic of the cell complex either, i.e.
\bea
\Delta \chi=\Delta V-\Delta E+\Delta F=0-1+1=0.
\eea 

In the case of the NGF model, for $d=2$ at each time we add a single $2$-dimensional cell glued to a single link. 
If the added cell is a triangle, then this adds a single node, $\Delta V=1$,  two links,  $\Delta E=2$, and one triangle,  $\Delta F=1$,  leading to 
\bea
\Delta \chi=\Delta V-\Delta E+\Delta F=1-2+1=0.
\eea
If, on the contrary, the added $2$-dimensional cell is a square, then it adds two new nodes, $\Delta V=2$,  three links, $\Delta E=3$, and one square  $\Delta F=1$,
 leading to 
\bea
\Delta \chi=\Delta V-\Delta E+\Delta F=2-3+1=0.
\eea
By similar direct inspection it can be easily shown that also for dimensions $d>2$ the topological moves that define the NGF evolution do not change the Euler characteristic.
The changes of $\Delta V,\Delta E,\Delta F$ and $\Delta \chi$ for all the discussed topological moves are summarized in Table $\ref{tab:euler}$.

\begin{table}
\begin{tabular}{|c|c|c|c|c|c|c|}
\hline
\multicolumn{2}{|c|}{Model} & Process & $+\Delta V$ & $-\Delta E$ & $+\Delta F$ & $\Delta \chi$ \\ \hline
\multirow{4}{*}{STC} & \multirow{2}{*}{$m=2$} & Triangle & $1$ & $-2$ & $1$ & $0$ \\ \cline{3-7}
&& Square & $1$ & $-2$ & $1$ & $0$ \\ \cline{2-7}
&\multirow{2}{*}{$m>2$} & $1$\textsuperscript{st} link & $1$ & $-2$ & $1$ & \multirow{2}{*}{$0$} \\ 
&& following links & $0$ & $-1$ & $1$ &  \\ \hline
\multirow{2}{*}{NGF} & \multirow{2}{*}{$d=2$} & Simplex & $1$ & $-2$ & $1$ & $0$ \\ \cline{3-7}
& & Orthoplex & $2$ & $-3$ & $1$ & $0$ \\ \hline
\end{tabular}
\caption{Changes in the number of nodes $\Delta V$, number of links $\Delta E$ and number of $2$-dimensional cells $\Delta F$ and corresponding change in the  Euler characteristic $\delta \chi$ for each topological move determining the evolution of the  STC and NGF high-order networks of dimension $2$. \label{tab:euler}}
\end{table}

\subsection{Emergent community structure}

\textit{Triadic closure} was recently proposed as a general, unifying mechanism to generate a community structure  \cite{bianconi2014triadic}. 
Such a process is frequently observed in social networks, where  open triads are often closed over time, and the density of triangles is remarkably high  \cite{newman2003social,  toivonen2006model, foster2011clustering}.
Models of network growth based on simple triadic closure have been shown \cite{bianconi2014triadic} to naturally lead to the emergence of community structure, provided that the network is sufficiently sparse.

\begin{figure}
\centering 
\includegraphics[width=1\columnwidth]{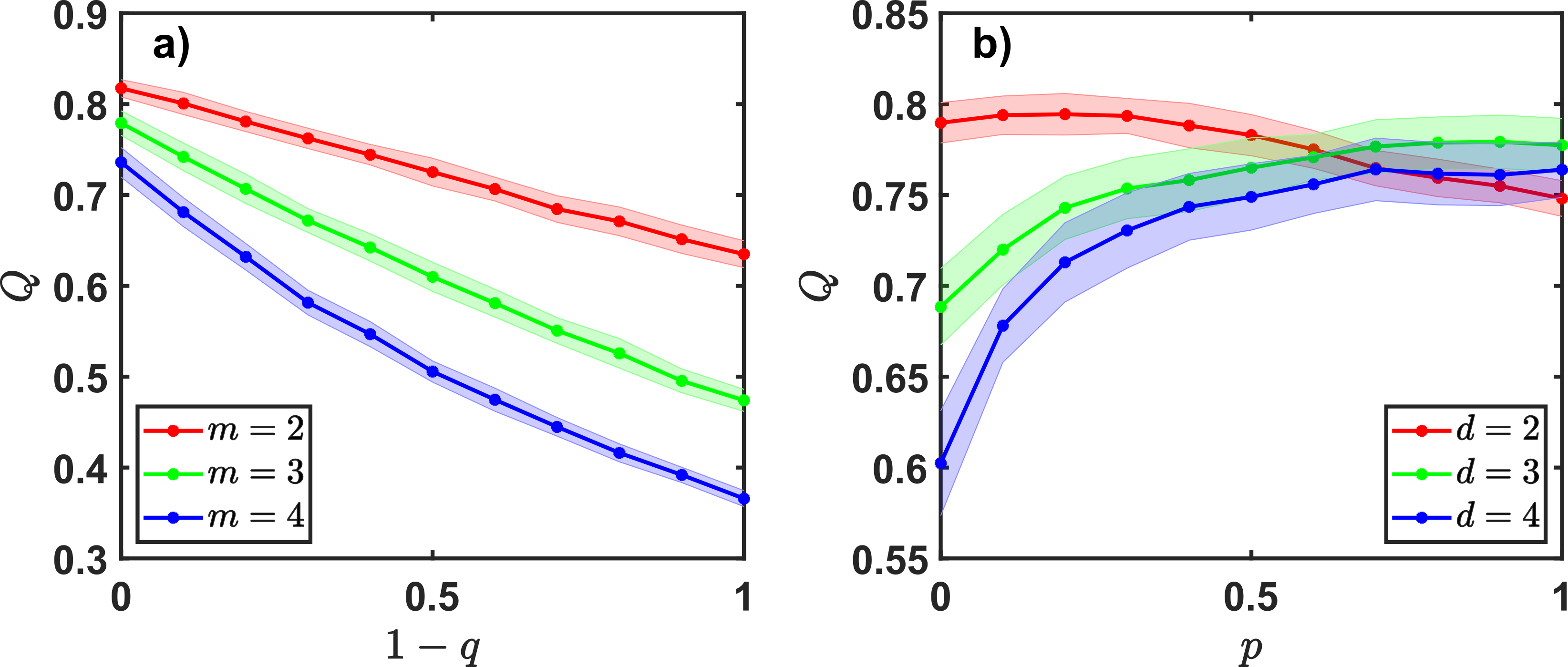}
\caption{Modularity $Q$ of the STC (panel a) and NGF (panel b) networks for different values of $m$ and $d$. Results are for $N=10^3$. Shaded areas indicate margin errors and are given by the standard deviation; results are averaged over $100$ network realizations. 
 \label{fig:fig2}}
\end{figure}

Interestingly, the non-equilibrium mechanisms leading to the emergence of a community structure are at work both in the STC and NGF models via the local topological moves.
Consequently, both models lead to network skeletons with a strong community structure, as indicated by the high values of the modularity coefficient $Q$ (see  Fig.\,$\ref{fig:fig2}$).
However, as $p$ increases the models display radically different behaviors: for STC networks $Q$ decays almost linearly with $p=1-q$, whereas for NGF networks with $d=3,\ 4$ it grows. 
In the STC model, as longer-range connections become more prominent, interconnections between regions also grow, decreasing the modularity. 
By contrast, orthoplexes in the NGF model are less interconnected with the rest of the network, as they only share one face with it, but have more faces than the simplices, leading to an increase in the modularity. In this perspective, the NGF model with $d=2$ stands out as its behavior is more complex: the modularity is relatively high for all $p$, following an inverse U-shape. In any case, the value of the modularity of the NGF model remains always significantly high, in contrast to the STC model. 

\section{Emergent geometrical network properties}

\subsection{Infinite Hausdorff dimension}

In this section we discuss the emergent geometrical properties of the STC and the NGF networks. 
An important geometrical notion that applies to network models with variable number of nodes, including regular lattices, is that of the Hausdorff dimension. 
The Hausforff dimension describes how the mean distance of the network $\ell$ scales with the total number of nodes (or network size) $N$ as $N$ goes to infinity.
$\ell$ is defined as
\begin{equation}
\ell = \frac{1}{N(N-1)} \sum_i \sum_{j\neq i} d_{ij},
\end{equation}
where $d_{ij}$ is the length of the minimum path between nodes $i$ and $j$. 
For regular lattices, the mean distance on the network scales as a power of the network size for $N\gg1 $, 
\bea
\ell \sim N^{1/d_H},\eea
where $d_H$ is the {\em Hausdorff dimension}.
In most random graphs, like Erd\"os-R\'enyi ones, $\ell$ scales logarithmically with the size  $N$, i.e.  $
\ell \sim \log N$, meaning that distances on the network are short, and it is possible to go from one node to any other passing through a  small number of intermediate vertices. 
This is the so-called \textit{small-world} property, which corresponds to an infinite Hausdorff dimension \bea
 d_H=\infty.
 \eea

\begin{figure}
\centering 
\includegraphics[width=1\columnwidth]{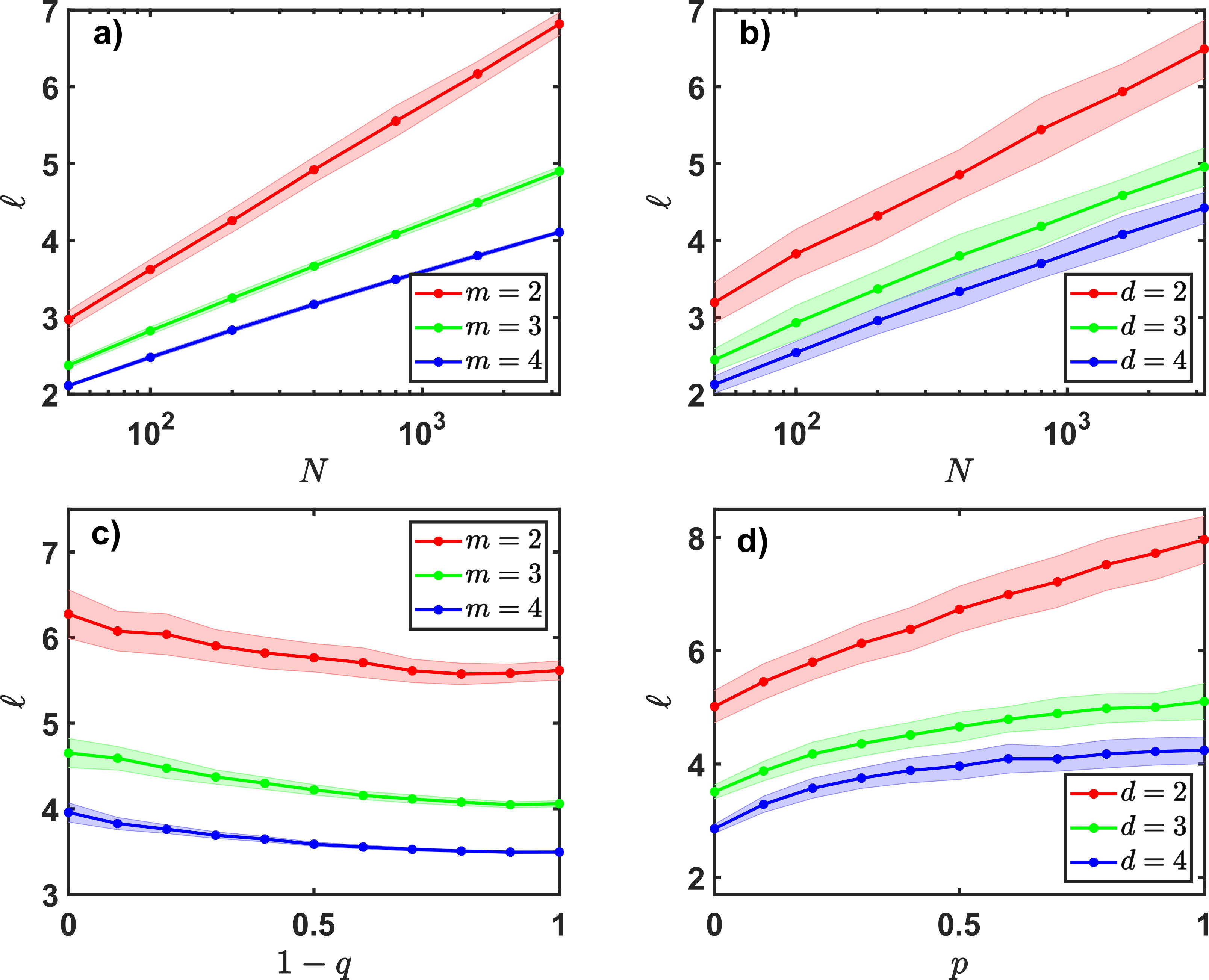}  
\caption{
We represent the  mean minimum path $\ell$  for the STC (panel a) and NGF (panel b) networks as a function of the network size $N$ for $p=0.5$.
$\ell$ grows logarithmically with $N$, which is indicative of the small-world property.
Panels c and d show $\ell$ for different values of $m$ and $d$, respectively for the STC and NGF models, for networks with $N=10^3$ nodes.
Results in all panels have been averaged over a set of $100$ network realizations, and the shaded areas indicate the error margins as given by the standard deviation.
\label{fig:fig3}}
\end{figure}

Both the STC and NGF models are small-world and have an infinite Hausdorff dimension, as shown in Fig. $\ref{fig:fig3}$a,b. 
However, for a fixed value of $N$, the characteristic distance $\ell=\ell (N)$ behaves remarkably differently in the STC and NGF models as a function respectively of $q$ and $p$ as it is shown in Fig. $\ref{fig:fig3}$c,d. 
In fact, in the STC model, as $p=1-q$  increases and we add more squares, distances in the network decrease as longer distance links become more common. 
This is evidenced by the decrease in the characteristic distance $\ell$ with $p=1-q$ (see Fig. $\ref{fig:fig3}$c).  
On the contrary, distances on the NGFs networks  increase as more orthoplexes are added (with increasing $p$), for any dimension $d$ (see Fig. $\ref{fig:fig3}$d). 
In order to better interpret this finding, we note that orthoplexes have more faces than simplices, and they are only glued to the existing network through one of them, \textit{de facto} increasing distances on the network.

\subsection{$\delta$-hyperbolicity of the STC and the NGF models}

\begin{figure}
\centering 
\includegraphics[width=0.8\columnwidth]{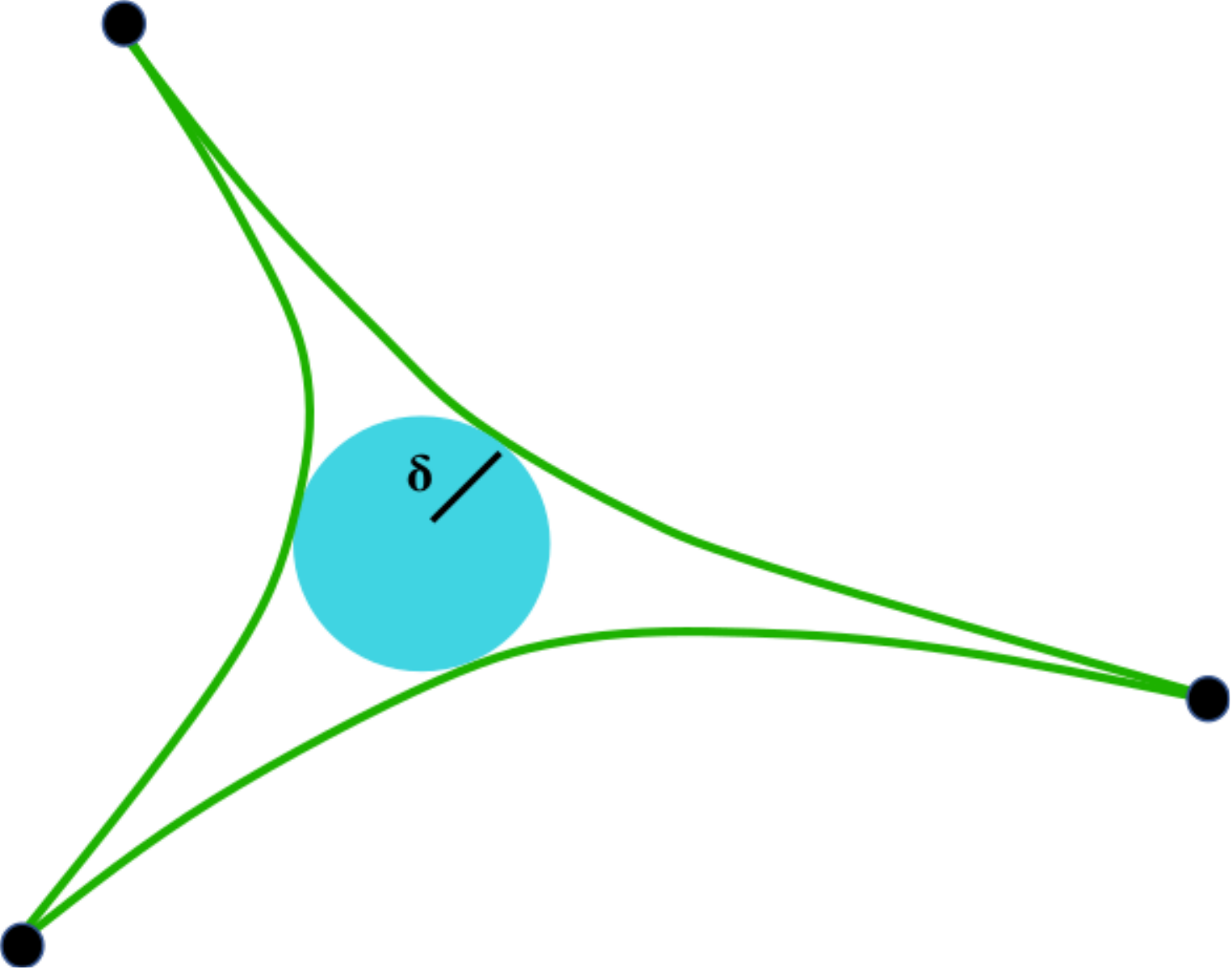}
\caption{ Schematic representation of a $\delta$-thin triangle formed by three nodes of the network and the shortest paths between the nodes. 
 \label{fig:fig4}}
\end{figure}

Hyperbolicity is an important geometrical aspect of real-world networks, and it is know to affects the efficiency of search  \cite{kleinberg2007geographic}, the efficacy of embedding algorithms  \cite{chami2019hyperbolic, chami2020low} and  the behavior of dynamical processes such as percolation  \cite{kryven2019renormalization}.

An important principle to test the hyperbolicity of real networks was proposed by Gromov  \citep{gromov1987hyperbolic}, who formulated the concept of $\delta$-hyperbolicity of networks  \cite{albert2014topological, narayan2011large, tadic1, vsuvakov2018hidden}. 
Hyperbolic spaces are characterized by having thin triangles such that the sum of their angles is less than $\pi$ (schematically represented in Fig. $\ref{fig:fig4}$).
The $\delta$-hyperbolicity measures how far a network is from an hyperbolic tree (which is $\delta=0$ hyperbolic)  and  quantifies how thin the triangles in the network are.
In particular, one can consider three nodes of the network and evaluate the minimum distance $\delta$ between the shortest paths connecting the three nodes, as illustrated in Fig. $\ref{fig:fig4}$.
If $\delta=0$ for every triangle of the network, then the network is a  tree. 
Interestingly, in a number of real networks $\delta$ remains always small, and in particular much smaller than the network diameter, indicating the hyperbolic geometry of the network.

\begin{figure}
\centering
\includegraphics[width=1\columnwidth]{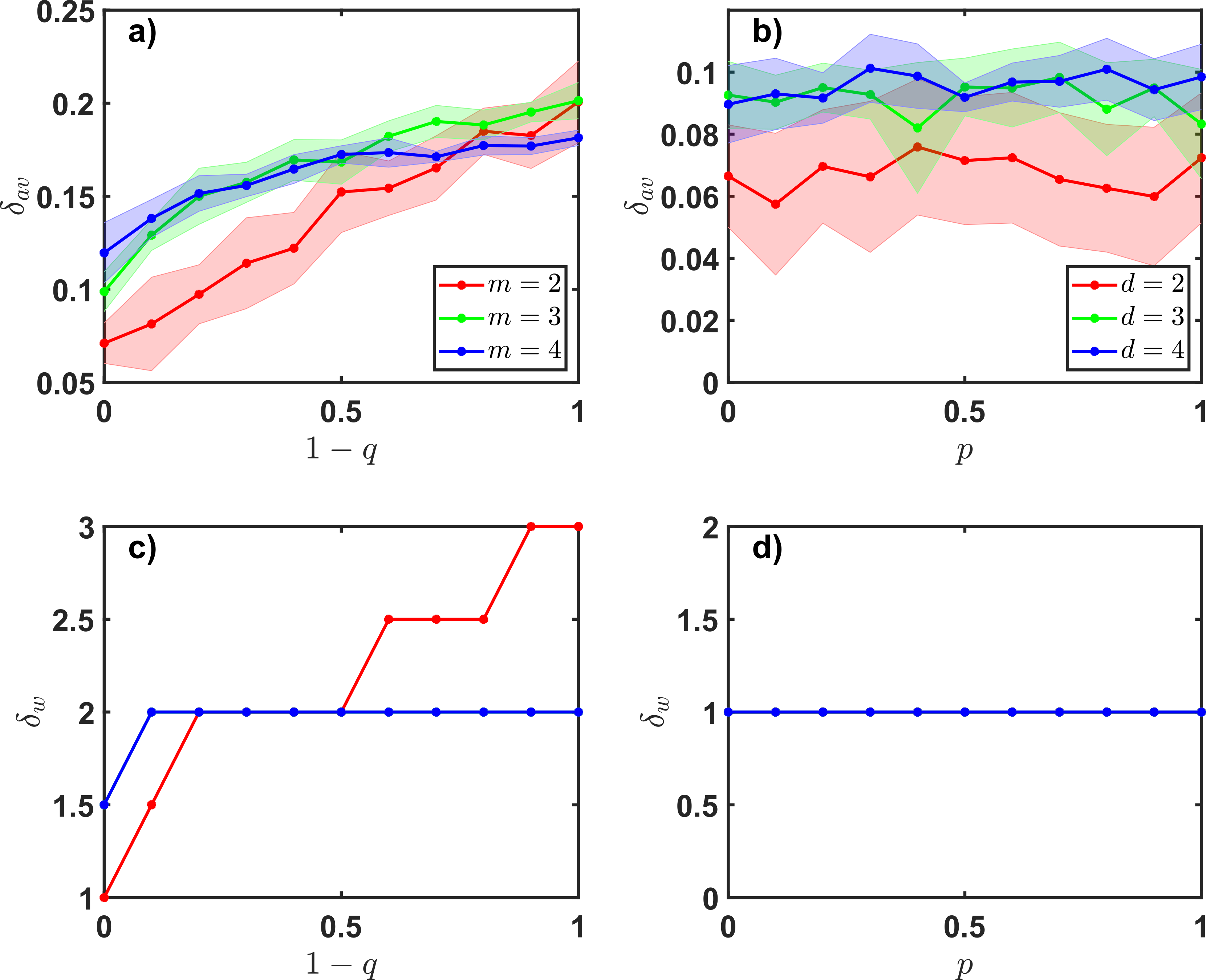} 
\caption{ $\delta$-hyperbolicity as given by the average $\delta_{av}$  for STC (panel a) NGF (panel b) networks and by the worst value $\delta_w$ (panel c for the STC networks and d for the NGF networks).
 \label{fig:fig5}}
\end{figure}

Here, in order to characterize the $\delta$-hyperbolicity of the STC and the NGF model we adopt the so-called four-point criterion  \cite{albert2014topological}.
We consider any quadruple of distinct nodes $(i_1,i_2,i_3,i_4)$ of the networks, and we choose their permutation   $(u_1,u_2,u_3,u_4) = i_{\pi(1)},i_{\pi(2)},i_{\pi(3)},i_{\pi(4)})$ such that the following inequalities hold,
\bea
S_{u_1,u_2,u_3,u_4}&\leq &M_{u_1,u_2,u_3,u_4},\nonumber \\
M_{u_1,u_2,u_3,u_4}&\leq& L_{u_1,u_2,u_3,u_4},
\label{eq:dis_delta}
\eea
where $S_{u_1,u_2,u_3,u_4},M_{u_1,i_2,u_3,u_4}$ and $L_{u_1,i_2,u_3,u_4}$ are defined as
\bea
S_{u_1,u_2,u_3,u_4}&=&d_{u_1,u_2}+d_{u_3,u_4},\nonumber \\
M_{u_1,u_2,u_3,u_4}&=&d_{u_1,u_3}+d_{u_2,u_4},\nonumber \\
L_{u_1,u_2,u_3,u_4}&=&d_{u_1,u_4}+d_{u_2,u_3}.
\eea
For any quadruple of nodes $(i_1,i_2,i_3,i_4)$, once we have found their permutation $(u_1,u_2,u_3,u_4)$ satisfying Eqs. ($\ref{eq:dis_delta}$) we put
\bea
\delta_{i_1,i_2,i_3,i_4}^{+}=\frac{1}{2}\left[L_{u_1,u_2,u_3,u_4}-M_{u_1,u_2,u_3,u_4}\right].
\eea
In order to evaluate the $\delta$-hyperbolicity of the network, we consider two metrics: the $\delta_{w}$ or worst (largest) value of $\delta^+$, and the average value of $\delta^+$, $\delta_{av}$, which are defined as 
\bea
\delta_w&=&\max_{(i_1,i_2,i_3,i_4)}\delta_{i_1,i_2,i_3,i_4}^{+},\nonumber \\
\delta_{av}&=&\left[\left(\begin{array}{c} N\\ 4\end{array}\right)\right]^{-1} \sum_{i_1,i_2,i_3,i_4}\delta_{i_1,i_2,i_3,i_4}^{+}.
\eea
Both the STC and the NGF networks are $\delta$-hyperbolic, as shown in Fig. $\ref{fig:fig5}$. 
In particular, the NGF has a constant value of  $\delta_w=1$, i.e. it can be interpreted as a ``tree of polytopes", whereas the STC admits a larger value of $\delta_w$, but this is always bounded by a finite constant as the network size increases. 
These results confirm that both the STC and NGF (with flavor $s=0$) models define emergent hyperbolic network geometries, even tough these are not discrete manifolds.
Note that, while the STC remains $\delta$-hyperbolic for any value of the parameter $q$, the model in Ref. \cite{bianconi2014triadic} allowing each new node to connect with  non-zero probability to  two or more nodes chosen randomly in the network is not $\delta$-hyperbolic.

\subsection{Area and Volume of STC and NGF models}

The ratio between the area $A$ and the volume $V$ of the cell-complexes generated by the STC and NGF model NGF models is also a very notable geometrical property of the considered models which is a further indication of their hyperbolicity.

\begin{figure}
\centering
\includegraphics[width=0.9\columnwidth]{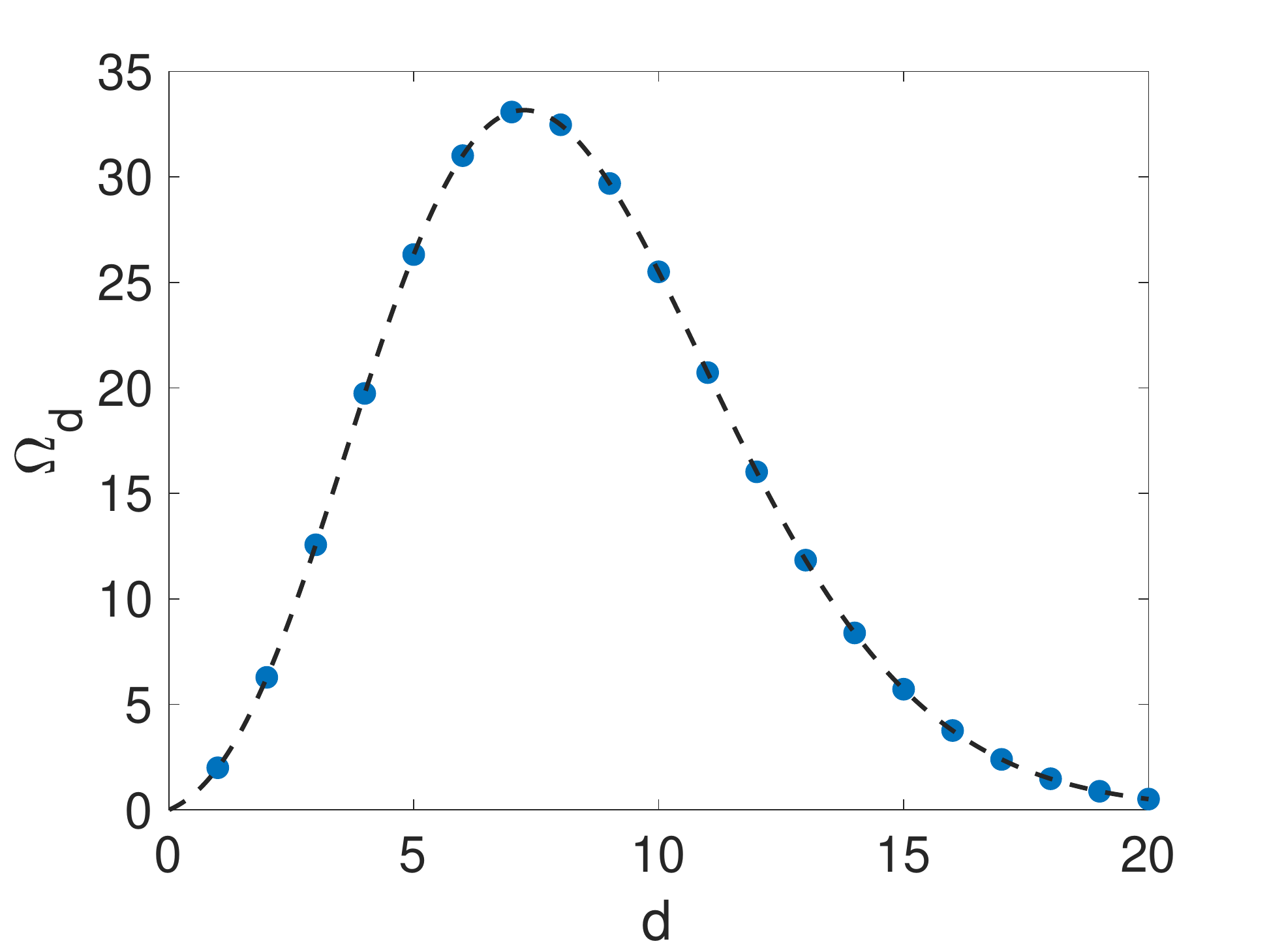}
\caption{ The area $A=\Omega_d$ of a unit ball in $d$ dimension is plotted versus $d$ and is shown to display a non-monotonous bahavior with a maximum for $d^{\dag}\simeq 7.25$.
 \label{fig:fig6}}
\end{figure}

Even tough the considered STC and NGF models do not define discrete manifolds, when investigating the properties of the ratio between their area $A$ and their volume $V$ it is useful to  refer first to known results valid for  Euclidean and hyperbolic manifolds.
For Euclidean balls of radius $R$, the area $A$ and the volume $V$ are given by  \cite{huang2009introduction}
\bea
A= \Omega_{d}R^{d-1},\quad
V= \frac{\Omega_d}{d} R^d,
\eea
where
\bea \label{omega}
\Omega_d=\frac{2\pi^{d/2}}{\Gamma(d/2)}.
\eea
Therefore, the area and volume of a ball with unitary radius, $A=\Omega_d$ and $V=\Omega_d/d$,
depend non-trivially on the dimension of the ball $d$.
In particular, they have a non-monotonic behavior, with the volume having a maximum for $d^{\star}\simeq 5.25$ and the area having a maximum for $d^{\dag}\simeq 7.25$ (see Fig. $\ref{fig:fig6}$).
Therefore, for high dimension $d$ both the area and the volume of the unit ball decrease with $d$.
More in general, the area-volume ratio of a ball of radius $R$ is given by
\bea
\frac{A}{V}=\frac{d}{R}.
\eea
Therefore, the ratio $A/V$ vanishes to zero in the limit of a ball of large radius, i.e. $A/V\to 0$ for $R\to\infty$.

A different behavior is observed for hyperbolic manifolds.
For $\mathbb{H}^d$ hyperbolic manifolds in  dimension $d$  \cite{ratcliffe1994foundations},  
a ball of radius $R$ in $\mathbb{H}^d$  has area and volume given by 
\bea
A&=&\Omega_d \sinh^{d-1}(R)\nonumber \\
V&=&\Omega_d\int_0^{R}\sinh^{d-1}(x)dx
\eea
with $\Omega_d$ given always by Eq. ($\ref{omega}$).
Thus, the scaling with $R$ changes and in this case the  area-volume ratio remains finite  in the $R\to \infty$ limit, i.e.
\bea
\lim_{R\to \infty}\frac{A}{V}=d-1.
\eea

Let us now explore the area-to-volume ration in the STC and the NGF models.
We define the area $A$ of a $d$-dimensional cell complex which is pure, (i.e. whose facets are all polytopes  of dimension $d$)
as given by the number of $(d-1)$-dimensional faces $\alpha$ that are incident to a single polytope (or equivalently with incidence number $n_{\alpha}=0$).
Similarly,  we define the volume $V$ of a $d$-dimensional cell complex which is pure,  as  the total number its $(d-1)$-dimensional faces.

Since the STC model is a $d=2$ dimensional cell complex for every value of $m$, its area $A$ is given by the number of links which are incident either to a single triangle or to a single square. 
The volume $V$ of STC is given by the total number of links.
For the NGF model in dimension $d$, the area $A$ is the number of $(d-1)$-simplices of the cell complex that are incident to a single $d$-dimensional cell (either a $d$-dimensional orthoplex or a $d$-dimensional simplex). 
The volume $V$ of the NGF is given by the total number of $(d-1)$-simplices.

\begin{figure}
\centering 
\includegraphics[width=1\columnwidth]{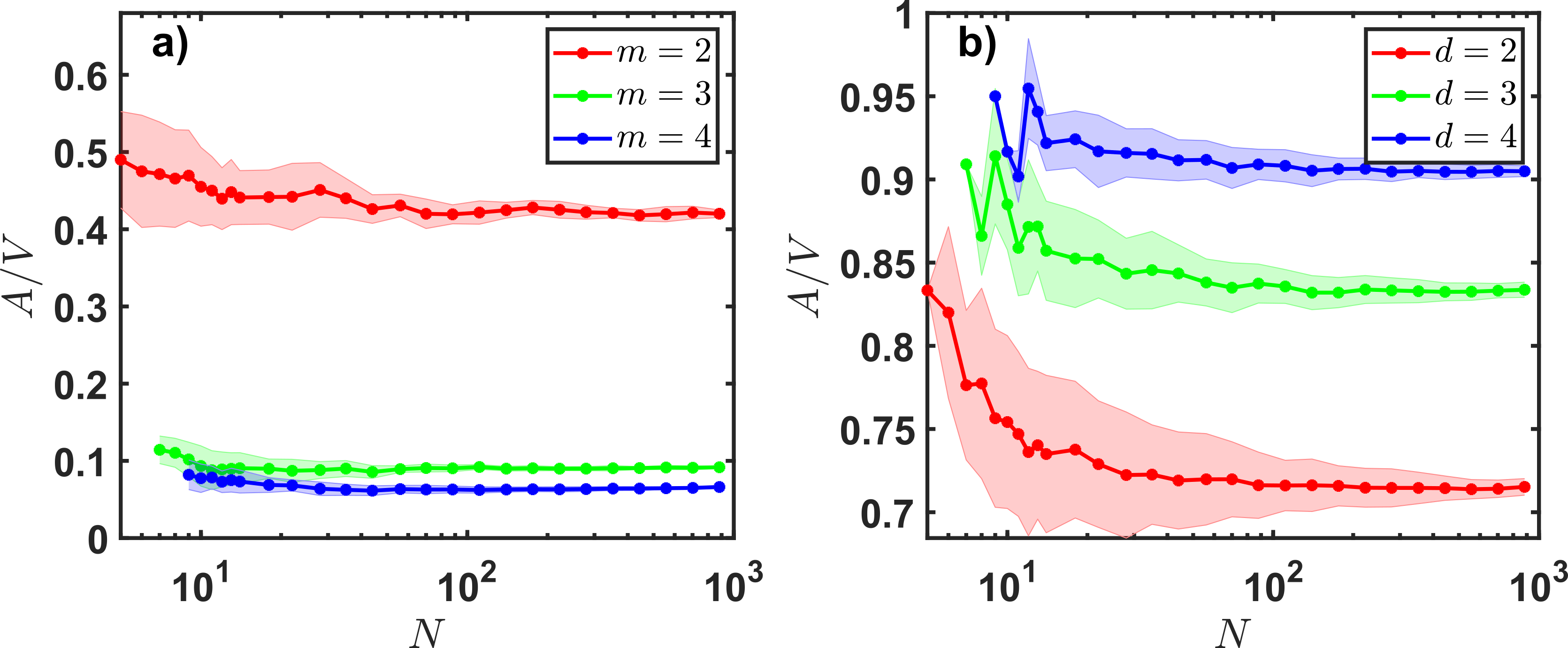}  
\caption{
We illustrate the convergence of $A/V$ during network growth 
for the STC (panel a ) and NGF (panel b) models, for different values of $m$ and $d$, and for $q=p=0.5$.  In all panels, results are averaged over $100$ network realizations, and the shaded areas indicate the error margins as given by the standard deviation. 
}
\label{fig:fig7}
\end{figure}

We are now in place to study the dependence with the network size of the ratio $A/V$ for the STC and  NGF models.
As it can be seen from  Fig. $\ref{fig:fig7}$, the ratio $A/V$ reaches  a constant value in the large network limit, confirming the hyperbolic nature of the models.
As shown in Fig. $\ref{fig:fig9}$c,d, the limiting value of the ratio $A/V$ depends on the value of $q$ (for the STC model) and $p$ (for the NGF model), even tough the dimension $d$ of the cell complex is independent of $q$ and $p$. 
Therefore, the ratio $A/V$ does not give an indication of the dimension $d$. In the next section we will show how the value of $A/V$ can be related instead to a different notion of dimension, the spectral dimension of the network. 

\section{Spectral dimension and diffusion dynamics}

\begin{figure*}
\centering 
\includegraphics[width=1\textwidth]{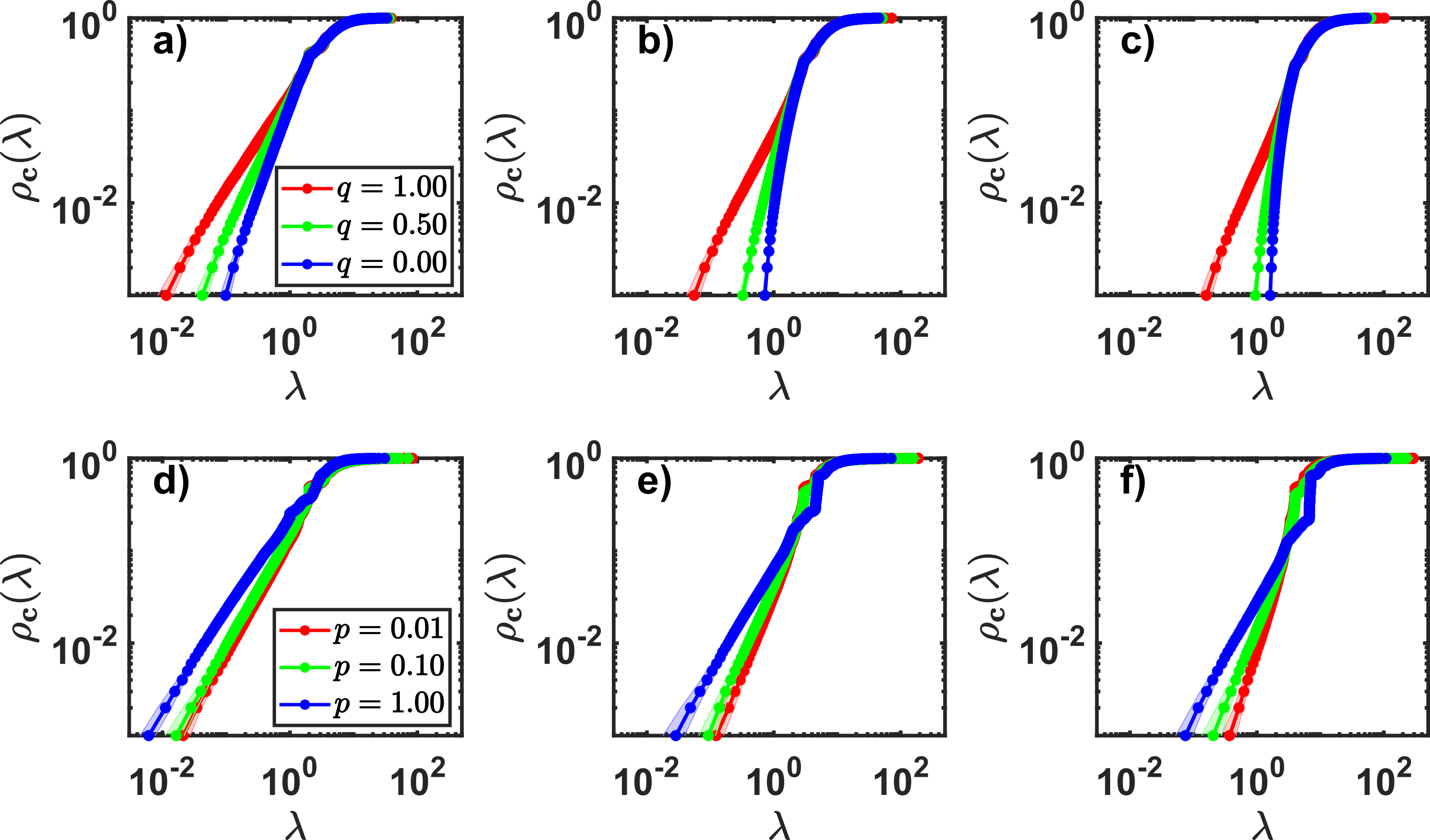}      
\caption{The cumulative density $\rho_c(\lambda)$ of eigenvalues $\lambda$, is shown for the STC (panels a,b,c, respectively for $m=2,\ 3,\ 4$) and the NGF models (panels d,e,f, respectively for $d=2,\ 3,\ 4$). 
We have used networks with  $N=10^3$ and parameters $p$ and $q$ shown in the legend. The data has  been averaged over $100$ network realizations. The error margins as indicated by the standard deviation are shown with shaded areas (although in most cases the error area overlaps with the data-points). 
\label{fig:fig8}}
\end{figure*}

Much information on the structure of a network is given by the properties of its associated Laplacian matrix \cite{rammal1984random, Burioni1996}.
For many complex networks, the Fiedler (second smallest) eigenvalue remains finite in the thermodynamic limit (the smallest eigenvalue is always zero by definition), and in such case the network is said to display a spectral gap. 
On the contrary, if the spectral gap closes as the system size grows, the network is said to have a finite spectral dimension  when the scaling of the cumulative density of eigenvalues of the Laplacian follows a power-law   \cite{Burioni1996, Burioni1999, hwang2010spectral}.

Here we consider the normalized Laplacian ${\bf L}$ with elements
\begin{equation} \label{eq:laplacian}
L_{ij} = \delta_{ij} - \frac{a_{ij}}{k_j},
\end{equation}
where $a_{ij}$ is the adjacency matrix of the network and $\delta_{ij}$ is the Kronecker's delta. 
${\bf L}$ has real non-negative eigenvalues $0=\lambda_1 \leq \lambda_2 \leq ... \leq \lambda_N$. 
The spectral density $\rho(\lambda)$ indicating the density of eigenvalues is defined as 
\bea
\rho(\lambda)=\frac{1}{N}\sum_{i=1}^N \delta(\lambda,\lambda_i),
\eea
where $\delta(x,y)$ indicates the delta function.
If the second smallest  eigenvalue $\lambda_2$ goes to zero in the large network limit, i.e. $\lambda_2\to 0 $ for $N\to \infty$, and the  density of eigenvalues $\rho(\lambda)$ for $\lambda \ll 1$ scales as 
\begin{equation}
\rho(\lambda) \simeq C\lambda^{d_S/2-1},
\end{equation}
with $C$ indicating a constant, we say that the network has  \emph{spectral dimension} $d_S$.
The spectral dimension can be interpreted as the dimension of the network as perceived by a random walker diffusing on it, and it is a notable feature of networks with a distinct geometrical nature. 
For Euclidean lattices in dimension $d$, the  spectral dimension coincides with the Hausdorff dimension, i.e. $d_S=d_H=d$.
However, in general networks the spectral dimension can strongly differ from the Hausdorff dimension   \cite{jonsson1998spectral,durhuus2007spectral}.

Interestingly, both the STC model and the NGF model display a finite spectral dimension $d_S\geq 2$ for most of their parameter values, which coexists with their infinite Hausdorff dimension $d_H=\infty$.
This spectral dimension can be tuned by changing the control parameters $q$ and $p$, respectively in the STC and in the NGF model. 
In particular, the spectral dimension of STC networks increases for smaller values of $q$, while $d_S$ decreases for NGF networks with larger values of $p$.

In order to provide evidence for this effect, in Fig. $\ref{fig:fig8}$  we show the cumulative spectral density $\rho_c(\lambda)$ for STC and NGF networks for a choice of the values $q$ and $p$, respectively.
In presence of a finite spectral dimension $d_S$, this cumulative density of eigenvalues should scale as a power-law for $\lambda\ll1$, i.e. 
\bea
\rho_c(\lambda)\simeq C' \lambda^{d_S/2},
\eea 
where $C'$ is a constant.
For $p=0$ ($q=1$), both STC and NGF networks have a finite spectral dimension, as evidenced by the power-law scaling of $\rho_c(\lambda)$.
For STC networks, as $q$ decreases $d_S$ increases, as evidenced by the increase in the slope of $\rho_c(\lambda)$, until for $q=0$, $\rho_c(\lambda)$ is no longer power-law distributed: the networks present a spectral gap. 
NGF networks, on the contrary, always have a finite spectral dimension. Moreover, $d_S$ decreases as $p$ is increased. 
The NGF networks also present some degenerate eigenvalues in the high portion of the spectrum,  due to the underlying network symmetries.

\begin{figure*}
\centering 
\includegraphics[width=0.9\textwidth]{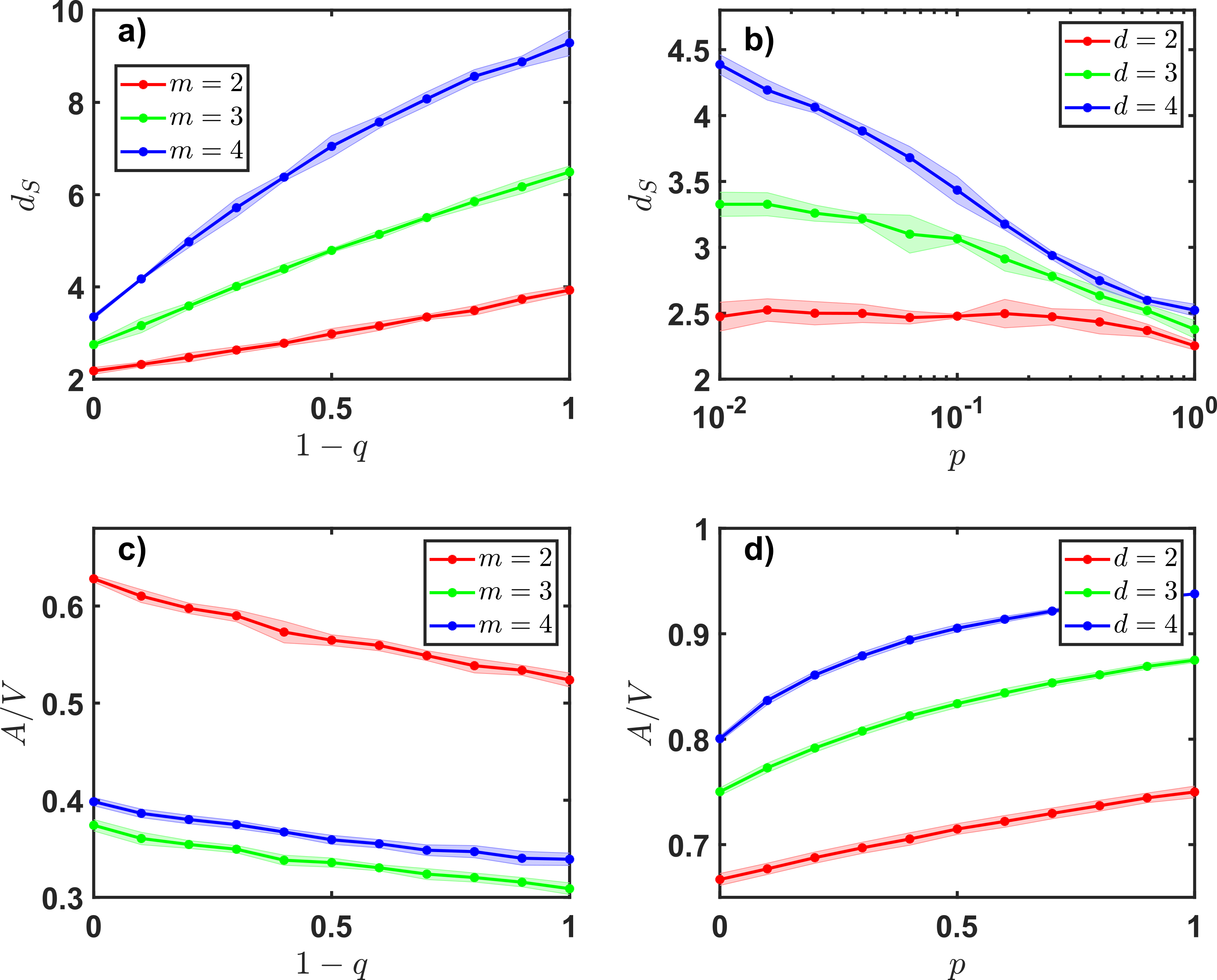}
\caption{We show the spectral dimension $d_S$ as a function of $p$ for STC ($m=2,\ 3,\ 4$, panel a, $p=1-q$) and NGF ($s=0$, $d=2,\ 3,\ 4$, panel b) networks. 
The values were measured from the $\rho_c(\lambda)$ curves as illustrated in Fig. $\ref{fig:fig8}$. 
Data form $400$ networks with $N=10^3$ was aggregated into $16$ groups of $25$ networks to construct average $\rho_c(\lambda)$ curves. 
$d_S$ was then measured independently for each group, and the average value is reported here together with the standard deviation between groups (as indicated by the shaded areas).
Panels c and d show
the area-volume ratio $A/V$ respectively for the STC and NGF models, for $m$ and $d$ as indicated in the legends. 
\label{fig:fig9}}
\end{figure*}

To better characterize this finding, we have measured $d_S$ in Fig. $\ref{fig:fig9}$a,b for STCs and NGFs of $m=2,\ 3\,  4$ and $d=2,\ 3,\ 4$, respectively. 
For STC networks, $d_S$ increases in an approximately linear manner with $p$. 
On the contrary, $d_S$ decreases in a non-linear manner for NGF networks. 
This result shows that, by changing the local topological moves by which the higher-order network skeleton evolves, it is possible to tune the corresponding value of the spectral dimension.
In particular, our results show that, in the considered hyperbolic higher-order network models, a smaller spectral dimension corresponds to a larger ratio $A/V$ between the area and the volume of the cell complex (see Fig. $\ref{fig:fig9}$c,d). 
Therefore, the ratio $A/V$ of the STC and NGF models is not indicative of their topological dimension, but rather it correlates with the spectral dimension $d_S$. 
This indicates that the different choice of topological moves used to generate the higher-order networks can at the same time change the area/volume ratio of the hyperbolic STC and NGF model and tune the value of their spectral dimension.

In order to illustrate how a different network structure and geometry affect the dynamical properties of the networks,  we have explicitly considered the diffusion dynamics of random walkers on the STC and NGF networks. 
Indeed, the return rate distribution $P_0(t)$ of random walkers diffusing on a network is directly linked to the spectral density by the equation  \cite{barrat2008dynamical}
\begin{equation}\label{eq:p0t}
P_0(t) = \int_0^\infty d\lambda e^{-\lambda t} \rho(\lambda).
\end{equation}
As a consequence, for networks with a finite spectral dimension,  the return distribution decays as a power-law whose exponent is determined by the spectral dimension, i.e.  
\bea
P_0(t) \propto t^{-d_S/2}
\eea for $t\ll 1$. 
This relation reveals the role of the  spectral dimension as a key spectral property, explicitly linking the structural and dynamical properties of a network. 

\begin{figure*}
\centering
\includegraphics[width=0.98\textwidth]{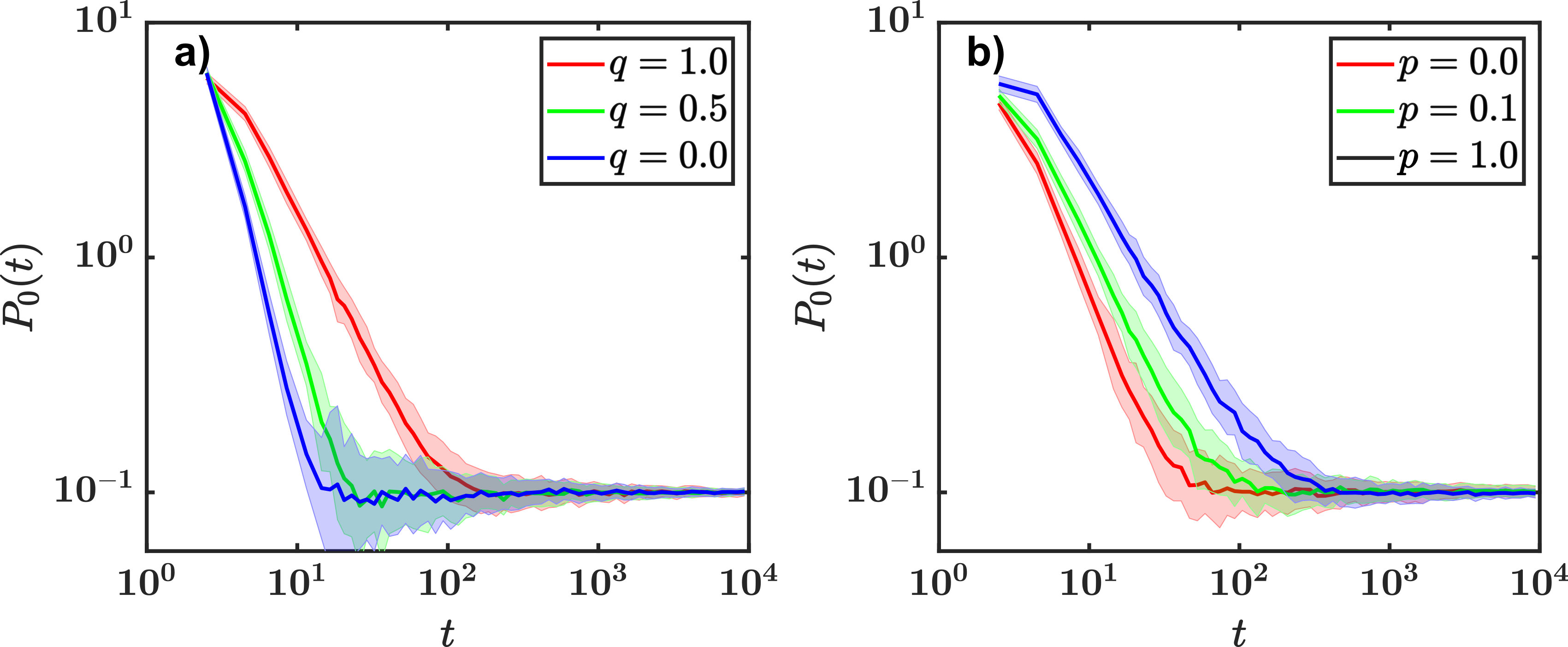} 
\caption{
Return-time probability $P_0(t)$ of the random walk for STC networks with $m=4$ (panel a) and NGFs with $d=4$, $s=0$ (panel b), for different values of $p$, as indicated by the legends, and $N=10^3$. Results are measured over $50$ realizations of the networks, and $20$ realizations of the RW dynamics for each network. 
Initially, $N_R=10^ 3$ walkers are placed on the network at randomly selected nodes $n_i$, $i=1,\ ...,\ N_R$. 
At each time step, each walker jumps with uniform probability to one of the networks of $n_i$, independently of the other walkers' positions. 
This process is iterated for $T_{RW} = 10^4$ steps, and the return probability $P_0(t)$ that a walker returns to its starting point after $t$ steps is measured. 
\label{fig:fig10}  }
\end{figure*}

By performing explicit simulation of the random walk dynamics, we  have measured  the return-time probability $P_0(t)$ for STC and NGF networks (see Fig.\,$\ref{fig:fig10}$).
 These results confirm  that the spectral dimension of the STC and the NGF models can be tuned by varying the parameter $q$ (for the STC model) and $p$ (for the NGF) model.   

\section{Conclusions}

Higher-order networks allow us to properly encode and investigate complex systems where interactions are not limited to two nodes at a time. In particular, their characterization beyond traditional metrics from statistical mechanics and network science reveals new features of the complex interplay between network structure and dynamics. 

In this work we proposed a general non-equilibrium framework which makes possible to obtain tunable emergent hyperbolic network geometries, and provide new insights on how topology and geometry affect diffusion dynamics. 
We introduced two models, namely the STC and NGF higher-order models, which are generated by iteration of simple, local, topological moves. 
We investigated how variations in these local rules are reflected in the geometrical properties of the higher-order networks. 
In particular, despite leaving the topological invariants of the higher-order network unchanged, we showed that local moves have the ability to modify the emerging geometrical and diffusion properties. 
We measured the diffusion properties of the considered models in terms of their spectral dimension,  a remarkable geometrical property of their network skeleton that  determines the return time distribution of a random walk crawling on it. 
We found that the spectral dimension does not display a universal value on the considered models, but on the contrary it can be tuned by modulating the ratio between the area and the volume of the higher-order models, explicitly governed by the choice of the topological moves. 
In particular, as the area-volume ratio grows, the spectral dimension of the model increases as well. The considered models of emergent network geometries  are also found to show other non-trivial features of real-world systems, including the small-world property, $\delta$-hyperbolicity and significant community structure.
We believe that our work reveals a new link between the geometry of a network and the properties of diffusion processes taking place on top of it, contributing at a fundamental level to a better understanding of the complex interplay between network structure and dynamics.

\section*{Acknowledgments}

Ana P. Mill\'an is supported by ZonMw and the Dutch Epilepsy Foundation, project number $95105006$. 
F.B. acknowledges partial support from the ERC Synergy Grant 810115 (DYNASNET). 
This work is  also supported by the Deutsche Forschungsgemeinschaft (DFG, German Research Foundation, N.D.) under Germany's Excellence Strategy EXC2181/1-390900948 (the Heidelberg STRUCTURES Excellence Cluster) and by the  Royal Society (IEC\textbackslash NSFC\textbackslash191147; G.B.).

\bibliographystyle{apsrev_titles}
\bibliography{bibliography}

\end{document}